\documentclass[twocolumn,showpacs,preprintnumbers,nofootinbib]{revtex4}
\usepackage{graphicx}

\usepackage{epstopdf}
\def\lsim{\mathrel{\rlap{\lower4pt\hbox{\hskip1pt$\sim$}}
    \raise1pt\hbox{$<$}}}
\def\gsim{\mathrel{\rlap{\lower4pt\hbox{\hskip1pt$\sim$}}
    \raise1pt\hbox{$>$}}}
\def\sqr#1#2{{\vcenter{\vbox{\hrule height.#2pt
         \hbox{\vrule width.#2pt height#1pt \kern#1pt
         \vrule width.#2pt}
         \hrule height.#2pt}}}}

\def\beq{\begin{equation}}
\def\eeq{\end{equation}}
\def\beqa{\begin{eqnarray}}
\def\eeqa{\end{eqnarray}}
\begin{document}

\title{The electromagnetic coupling and the dark side of the Universe}

\author{M. C. Bento}
\altaffiliation[ ] {Also at Centro de F\'isica Te\'orica e de Part\'iculas, IST. Email
address: bento@sirius.ist.utl.pt}

\author{O. Bertolami}
\altaffiliation[ ] { Also at Centro de F\'isica dos Plasmas, IST. Email address:
orfeu@cosmos.ist.utl.pt}

\author{P. Torres}
\altaffiliation[] {Also at Centro de F\'isica Te\'orica e de Part\'iculas, IST. Email
address: paulo.torres@ist.utl.pt }

\affiliation{ Departamento de F\'isica, Instituto Superior T\'ecnico \\
Av. Rovisco Pais 1, 1049-001 Lisboa, Portugal}

\vskip 0.5cm

\date{\today}

\begin{abstract}

We examine the properties of dark energy and dark matter through the study of the
variation of the electromagnetic coupling. For concreteness, we consider the unification
model of dark energy and dark matter, the generalized Chaplygin gas model (GCG),
characterized by the equation of state 
$p=-\frac{A}{\rho^\alpha}$, where $p$ is the pressure, $\rho$ is the energy density and $A$ 
and $\alpha$ are positive constants. The coupling of electromagnetism with the GCG's scalar 
field can give rise to such a variation. We compare our results with experimental data, and 
find that the degeneracy on  parameters $\alpha$ and $A_s$, $A_s \equiv A / \rho_{ch0}^{1+\alpha}$, 
is considerable.

\vskip 0.5cm

\end{abstract}

 \pacs{98.80.-k,98.80.Cq,12.60.-i\hspace{4cm} Preprint DF/IST-3.2006}

\maketitle
\vskip 2pc
\section{Introduction}

It is well known that, if four dimensional physics arises from higher dimensional
theories, then the presumably fundamental constants of the four dimensional theory are
actually time dependent. It is therefore particularly relevant to search for these
variations and for possible correlations with striking features of the Universe evolution.
Characterization of these correlations may, in turn, provide relevant hints on the nature
of the main constituents of the Universe, namely dark matter and dark energy.

Evidence suggesting that, for instance, the proton-electron mass ratio has varied has been
recently reported \cite{p-e}. This arises from comparison of the absorption spectra of
$H_2$ in the laboratory with that
 of two clouds of $H_2$ about $12\times 10^9$  light years away. Furthermore, the observation of the
 spectra of quasars (QSOs) seems to indicate a time-dependent fine structure parameter
 \cite{Murphy,Chand1,Chand2}.
Such observations lead to, for $0.2<z<3.7$ \cite{Murphy}, \beq
\frac{\Delta\alpha_{em}}{\alpha_{em}}=(-0.54\pm0.12)\times10^{-5}~, \label{murphy} \eeq
\noindent at $4,7 \sigma$.

More recent observations suggest, however, that for $0.4<z<2.3$ \cite{Chand1,Chand2},
\beq
\frac{\Delta\alpha_{em}}{\alpha_{em}}=(-0.06\pm0.06)\times10^{-5}~,
\label{chand}
\eeq
\noindent at $3 \sigma$.

The Oklo natural reactor also provides a bound on the variation of the electromagnetic coupling,
at $95 \%$ CL, for $z=0.14$ \cite{oklo1,oklo2,oklo3}
\beq
-0.9\times 10^{-7}<\frac{\Delta\alpha_{em}}{\alpha_{em}}<1.2\times 10^{-7}~.
\label{oklo}
\eeq

Estimates of the age of iron meteorites, corresponding to $z=0.45$, combined with a measurement
of the Os/Re ratio resulting from the radioactive decay $^{187}Re \to~^{187}Os$, yields \cite{met1,met2,met3}:
\beq
\frac{\Delta\alpha_{em}}{\alpha_{em}}=(-8\pm8)\times10^{-7}~,
\label{meteorites1}
\eeq
\noindent at $1 \sigma$, and
\beq
-24\times 10^{-7}<\frac{\Delta\alpha_{em}}{\alpha_{em}}<8\times 10^{-7}~,
\label{meteorites2}
\eeq
\noindent at $2 \sigma$.

Observations of the hyperfine frequencies of the $^{133}Cs$ and $^{87}Rb$ atoms in their electronic
ground state, using several laser cooled atomic fountain clocks, gives, at present ($z=0$):
\beq
{\left\vert \frac{\dot{\alpha} _{em}}{\alpha_{em}}\right\vert} < 4.2 \times
10^{-15}~\mbox{yr}^{-1}~,
\eeq
\noindent
 where the dot represents differentiation with respect to cosmic time. Stricter
 bounds arise from the measurement of the $1s-2s$ transition of the atomic hydrogen and
comparison with a previous measurement with respect to the ground
state hyperfine splitting in $^{133}$Cs and combination with the drift
of an optical transition frequency in $^{199}$Hg$^+$, which yields:
\beq
\frac{\dot{\alpha}_{em}}{\alpha_{em}} =
(-0.9 \pm 4.2) \times 10^{-15}~\mbox{yr}^{-1}~.
\eeq

As already pointed out, a spacetime dependence on the fine structure parameter arises
naturally in higher dimensional fundamental theories, as shown in several theoretical
investigations \cite{fund-th}. It is therefore relevant to deepen the experimental
observations and search for realistic models that allow for such variation.

Recent observations clearly indicate that the Universe is undergoing a period of accelerated expansion 
\cite{expansion}, which suggests that it is dominated by a form of energy density with negative pressure 
usually referred to as dark energy. The most obvious candidate to explain this accelerated expansion 
is an uncanceled cosmological constant \cite{BentoUncanceledCC}. This is, of course, problematic 
as it involves a serious fine-tuning difficulty. Other possible explanations include quintessence-type 
models \cite{quint} with one \cite{quint1} or two \cite{quint2} scalar fields, k-essence \cite{k-ess} 
and exotic equations of state like in the case of the Chaplygin gas and its generalized version \cite{GCG3,bil}.

In this context, it is relevant to question whether these two observational facts are related. This
issue has already been addressed in the context of quintessence models \cite{carroll}, $N=4$ 
supergravity models \cite{supergrav}, Dirac-Born-Infeld inspired models \cite{mohammad} and 
a Lorentz invariance violating model \cite{BM}.

Given that scalar fields are a common feature in most of the above mentioned approaches, 
one considers the coupling of such a scalar field
to electromagnetism and hence a variation of the fine structure parameter, $\alpha_{em}$, 
can arise \cite{Bekenstein}. The bounds expressed in Eqs.~(\ref{murphy})-(\ref{meteorites2}) 
allows one to place constraints on the coupling.

For sure, the standard model of particle physics does not fix the coupling between the scalar 
field and electromagnetism, hence it is relevant to analyze some realistic alternatives. In 
particular, we consider the generalized Chaplygin gas (GCG) since it is a promising 
possibility from the phenomenological point of view \cite {GCG6,realGCG,GCG1,GCG2,GCG4,GCG5,GCG7}.

This paper is organized as follows: In section II we review the GCG model and study
the coupling with electromagnetism. In section III we use experimental data to constrain 
the model. We present our conclusions in section IV.

\section{Description of the Chaplygin gas model}

The GCG model considers an exotic perfect fluid described by the equation of state
\cite{GCG3} 
\beq 
p = -\frac{A}{\rho ^{\alpha}_{ch}}~, 
\label{GCGes} 
\eeq 
\noindent 
where $A$ is a positive constant and $\alpha$ is a constant in the range $0\leq \alpha \leq 1$.
The covariant conservation of the energy-momentum tensor for an homogeneous and isotropic
spacetime implies that in terms of the scale factor, $a$: 
\beq 
\rho_{ch} = \Bigl[ A +
\frac{B}{a^{3(1+\alpha)}}\Bigr]^{\frac{1}{1+\alpha}}~. \label{GCGrho}
\eeq

\subsection{Real scalar field}

Following Ref. \cite{realGCG}, we describe the GCG through a real scalar field. We start with
the Lagrangian density
\beq
{\cal L}=\frac{1}{2}\dot\phi^2-V(\phi)~,
\label{realL}
\eeq
\noindent where the potential, for a flat Universe, has the form \cite{realGCG}
\beq
V=V_0e^{3(\alpha-1)\phi}\Bigl[cosh^{\frac{2}{\alpha+1}}\Bigl(\frac{m\phi}{2}\Bigr)
+cosh^{-\frac{2\alpha}{\alpha+1}}\Bigl(\frac{m\phi}{2}\Bigr)\Bigr]~,
\label{realV}
\eeq
\noindent where $V_0$ is a constant and $m=3(\alpha+1)$.

For the energy density of the field we have
\beq
\rho_\phi=\frac{1}{2}\dot\phi^2+V(\phi)=\Bigl[ A +
\frac{B}{a^{3(1+\alpha)}}\Bigr]^{\frac{1}{1+\alpha}}=\rho_{ch}~. \label{realRHO}
\eeq

Using Eq.~(\ref{GCGes}), we can write the pressure as follows
\beq
p_{\phi}=\frac{1}{2}\dot\phi^2-V(\phi)=-\frac{A}{\Bigl[ A + \frac{B}
{a^{3(1+\alpha)}}\Bigr]^{\frac{\alpha}{1+\alpha}}}~,
\label{realP}
\eeq

and then
\beq
\dot\phi^2=\frac{B}{a^{3(1+\alpha)}\Bigl[ A + \frac{B}{a^{3(1+\alpha)}}\Bigr]^{\frac{\alpha}{1+\alpha}}}~,
\label{realPHI1}
\eeq
\noindent
which, after integration, yields in terms of the redshift, once one assumes that at 
present, $a_0 =1$:

\beq
\phi (z)=\frac{1}{3(1+\alpha)}ln\left[\frac{\sqrt{\frac{A_s}{1-A_s}(1+z)^{-3(1+\alpha)}+1}-1}
{\sqrt{\frac{A_s}{1-A_s}(1+z)^{-3(1+\alpha)}+1}+1}\right]~.
\label{integratedPHI}
\eeq

It is easy to show that, at present

\beq
\phi_0=\frac{1}{3(1+\alpha)}ln\left(\frac{1-\sqrt{1-A_s}}{1+\sqrt{1-A_s}}\right)~.
\label{phi0}
\eeq


\subsection{Coupling with electromagnetism}

We consider now the interaction between the scalar and the electromagnetic field following a proposal
 by Bekenstein \cite{Bekenstein}

\beq
{\cal L}_{em} =\frac{1}{16\pi} f\left(\phi\right)F_{\mu\nu}F^{\mu\nu}~,
\eeq
\noindent
 where $f$ is an arbitrary function. Given that the variation of the
electromagnetic coupling is small, we expand this function up to first order
\beq
f\left(\phi\right)=\frac{1}{\alpha_{em 0}}\left[1+\xi \left(\phi-\phi_0\right)\right]~,
\eeq 
\noindent where $\xi$ is a constant. It follows that fine structure parameter, $\alpha_{em}$, is given by
\beq 
\alpha_{em} =\left[f\left(\phi\right)\right]^{-1}=\alpha_{em 0}\left[1-\xi\left(\phi-\phi_0\right)\right]~, \label{alpha2}
\eeq
\noindent
and hence, for its variation, we obtain
\beq \frac{\Delta\alpha_{em}}{\alpha_{em}}=\xi\left(\phi-\phi_0\right)~.
\label{delta2}
\eeq

The rate of variation of $\alpha$ at present is given by
\beq
\frac{\dot \alpha_{em}}{\alpha_{em}}= - \xi \frac{d\phi}{dy} H_0~,
\eeq
\noindent
where $y \equiv 1+z$ and $H_0=100 h~km~s^{-1} Mpc^{-1}$ is the Hubble constant.

We should take in account the Equivalence Principle experiments, which implies \cite{equivalence}
\beq
\xi \leq 7 \times 10^{-4}~,
\label{zetaF}
\eeq
\noindent where $\xi$ gives, at first order, the coupling between the scalar and electromagnetic fields.

\begin{figure}[t]
\includegraphics[width=7cm]{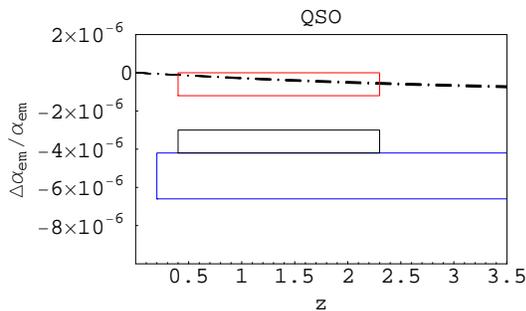}
\caption[fig:resultsQSO]{Evolution of $\frac{\Delta \alpha_{em}}{\alpha_{em}}$ for
$\xi=10^{-6}$,  $\alpha=0,~0.2,~0.4,~0.6,~0.8,~1$ (top to bottom). The lines overlap because of the
degeneracy on $\alpha$ (see Fig.\ref{fig:resultsNB}). The box corresponds to QSO bounds of
Refs. \cite{Chand1,Chand2} and \cite{Murphy} (top to bottom).}
 \label{fig:resultsQSO}
\end{figure}
\begin{figure}[t]
\includegraphics[width=7cm]{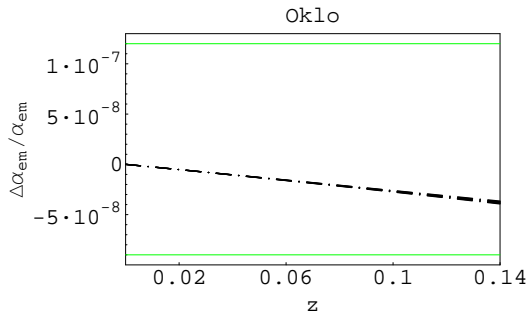}
\caption[fig:resultsOklo]{As for Figure \ref{fig:resultsQSO}. The box and the redshifts
correspond to Oklo bounds.} \label{fig:resultsOklo}
\end{figure}
\begin{figure}[t]
\includegraphics[width=7cm]{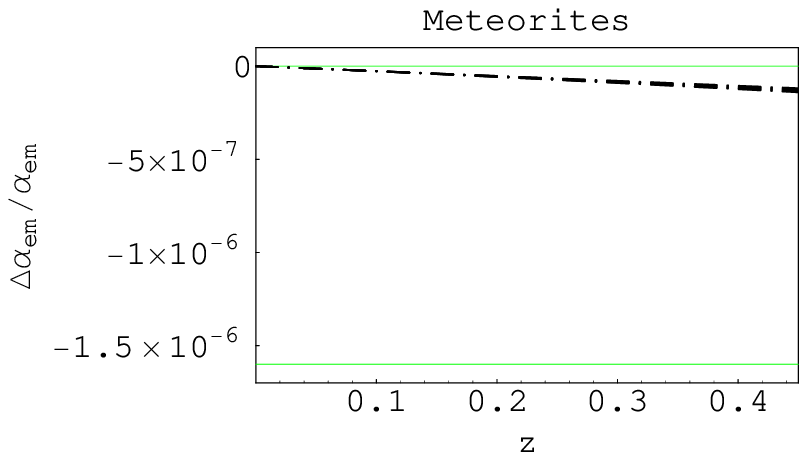}
\caption[fig:resultsMeteo]{As for Figure \ref{fig:resultsQSO}. The box and the redshifts
correspond to Metorites bounds.} \label{fig:resultsMeteo}
\end{figure}
\begin{figure}[t]
\includegraphics[width=7cm]{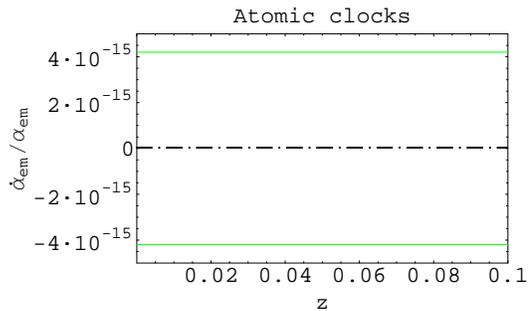}
\caption[fig:resultsAtomic]{As for Figure \ref{fig:resultsQSO}. The box and the redshifts
correspond to the bounds provided by atomic clocks ($\alpha=0,~0.2,~0.4,~0.6,~0.8,~1$; bottom to top)} \label{fig:resultsAtomic}
\end{figure}

\begin{figure*}[t]
\includegraphics[width=14cm]{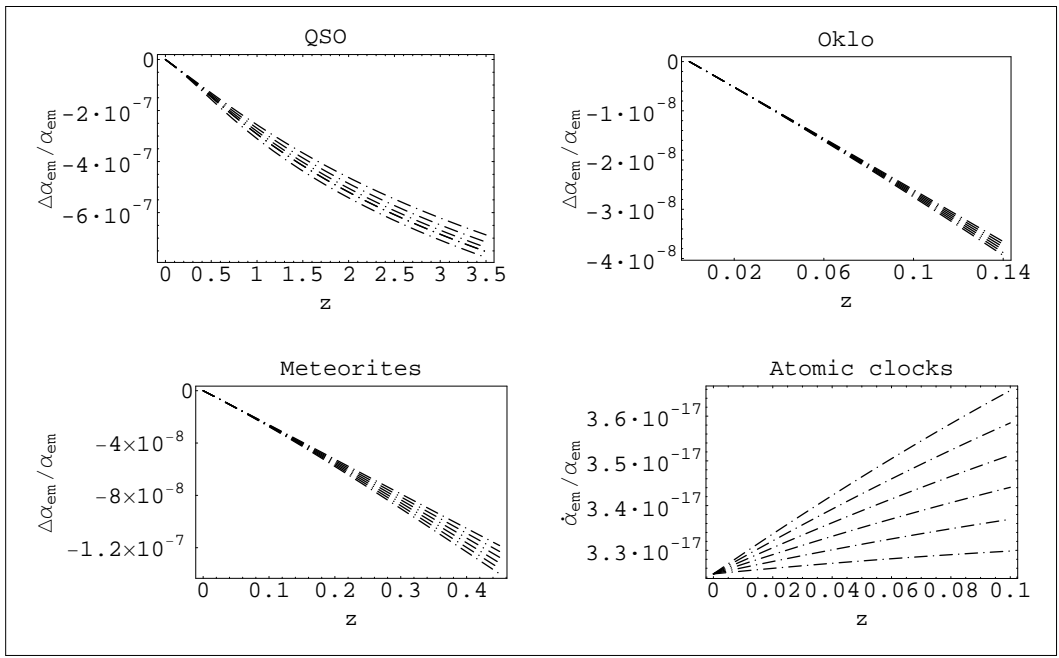}
\caption[fig:resultsNB]{As for Figures \ref{fig:resultsQSO}-\ref{fig:resultsAtomic}, but without the experimental bounds.}
\label{fig:resultsNB}
\end{figure*}

\section{Results}

We choose the Hubble constant, $h=0.71$, and adjust the coupling parameter, $\xi$, for
different values of $\alpha$ and $A_s$ in order to satisfy the bounds on the evolution of
$\alpha_{em}$.

Considering the set of parameters $A_s$ and $\alpha$, with $0.7 \leq A_s <1$ and for
$\xi\simeq 10^{-5}$, we find consistency with both meteorite and atomic clocks bounds.
However, the model does not fit QSO and Oklo data with this set of parameters. We have to
take $\xi\simeq 10^{-6}$ to obtain consistency with all bounds considered here. In this
case, however, the degeneracy on both parameters, $\alpha$ and $A_s$, is considerable,
that is, any values in the range $0\leq \alpha\leq 1$ and $0<A_s <1$ are consistent with
data. Figures \ref{fig:resultsQSO}, \ref{fig:resultsOklo}, \ref{fig:resultsMeteo} and
\ref{fig:resultsAtomic} summarize our results for $\xi=10^{-6}$ and $A_s=0.8$, for several
values of $\alpha$ in the range $0\leq \alpha\leq 1$.

\section{Discussion and Conclusions}

In this work, we have investigated the implications of the coupling to electromagnetism of
a scalar field model that describes in an unified fashion dark energy and dark matter, the
GCG model. The experimental bounds provided for such variation, namely QSOs, Oklo's
natural reactor, meteorite and atomic clocks, were used to constrain the model. We find
that the model is consistent with QSO, Oklo, meteorite and atomic clocks bounds
simultaneously for $\xi \simeq 10^{-6}$, for any value of the GCG's parameters $\alpha$
and $A_s$ in the ranges $0\leq \alpha \leq 1$ and $0<A_s<1$, respectively. Moreover, in
order to be consistent with the data, $\xi$ must be about three orders of magnitude
smaller than the upper bound implied by the Equivalence Principle (cf. Eq. (\ref{zetaF})).

\begin{acknowledgments}

The work of P.T. is supported by Funda\c{c}\~ao para a Ci\^encia e a Tecnologia (FCT, Portugal) under
the grant SFRH/BD/25592/2005.
The work of M.C.B. and O.B. is partially supported by the FCT project POCI/FIS/56093/2004.

\end{acknowledgments}

\end{document}